\documentclass{elsart5p}
\journal{Physics Letters A}

\usepackage{amsmath}
\usepackage{amsfonts}
\usepackage{amssymb}
\usepackage{graphicx}
\usepackage{epstopdf}

\begin{document}

\begin{frontmatter}
\title{Using reservoir computer to predict and prevent extreme events}

\author{Viktoras Pyragas and Kestutis Pyragas}
\address{Center for Physical Sciences and Technology, Sauletekio al. 3, LT-10257 Vilnius, Lithuania}

\begin{abstract}
We show that a reservoir computer is an effective tool for model-free prediction of extreme events in deterministic chaotic systems. This prediction allows us to suppress unwanted extreme events, by applying weak control perturbations to the system at times preceding expected extreme events. The effectiveness of such a prediction and prevention strategy is demonstrated for a system of globally coupled FitzHugh-Nagumo neurons and for a system of two almost identical unidirectionally coupled chaotic oscillators.      
\end{abstract}


\end{frontmatter}

\section{\label{sec1} Introduction}

The dynamics of various natural and engineering systems may exhibit rare, recurrent, and strong deviations from regular behavior. Such extreme events are of great interest  to various scientific disciplines \cite{Sergio2006}, since they often can have a serious impact on human life. Examples include oceanic rogue waves \cite{Disthe2008}, earthquakes \cite{Ohnaka2013}, shocks in power grids \cite{Crucitti2004}, stock market crashes \cite{Sornette2003}, and epileptic seizures \cite{Engel2007}. Thus, predicting and preventing extreme events is highly desirable. Extreme events often occur spontaneously without visible warning signs. They are rare in the sense that the frequency at which they occur is significantly less than the typical frequency of the system, and they are extreme in the sense that their amplitudes are several times higher than the standard deviation of the observed quantity.

Extreme events are analyzed in terms of both stochastic and deterministic models. Models of the first type usually have stable equilibria, and noise provides a transition between them, causing extreme events 
\cite{Forgoston2018}. Such events can be predicted only statistically. In deterministic models, extreme events are the result of the intrinsic nonlinear dynamics of the system. Even simple nonlinear systems can generate very complex chaotic dynamics, and determining the mechanism underlying extreme events is often a difficult task. Prediction of extreme events in deterministic systems can benefit from the fact that the current state of the system uniquely determines its future state, but it is limited by a sensitive dependence on the initial conditions. Forecasting strategies are usually based on the search for a precursor (indicator), which contains signs of early warning of upcoming extreme events. For low-dimensional systems, the precursor can be identified from a single scalar observable by using delay coordinate embedding techniques (see, e.g., \cite{Zaldivar2005}), while for high-dimensional systems,  knowledge of system equations may be required 
\cite{Farazmand2016}. A recent review of the mechanisms and prediction of extreme events can be found in Ref.~\cite{faraz18}.

In this paper, we appeal to a reservoir computer approach \cite{jaeger04,verst2007,Mantas09,Wyffels2010,Luk2012,Ott2017,Ott2017a,Ott2018,gaut2018,Ott2018b,Ott2018a}  to predict and prevent extreme events.  The approach employs a nonlinear input-output  neural network with randomly generated parameters, and uses linear regression to choose ``output weights'' that fit the network output to a set of ``training data.'' This approach is computationally simpler compared with other artificial neural network approaches, since only output weights are adjusted by the training process, while the  network parameters are fixed. In the pioneering work \cite{jaeger04} and recent publications \cite{Ott2017,Ott2017a,Ott2018,gaut2018,Ott2018b,Ott2018a,pyr2019} the reservoir computer (RC) has been successfully used to predict various low-dimensional and spatio-temporal chaotic systems. However, the previously discussed systems do not belong to the class of models generating extreme events. The prediction of extreme events is complicated by the fact that a local Lyapunov exponent, close to extreme events, may be significantly larger than the global (average) Lyapunov exponent of a chaotic attractor. Nevertheless, here we show that extreme events  can also be successfully predicted using a reservoir computer. We demonstrate this on two model systems generating extreme events. One of them is the system of globally coupled FitzHugh-Nagumo (FHN) neurons introduced and analyzed in Refs.~\cite{Feudel2013,Feudel2014}. We consider two variants of this model, consisting of a small and a large number of neurons. In this model, we introduce control variables and show that extreme events can be suppressed by applying small perturbations  at times preceding the predicted extreme events. As the second model, we consider unidirectionally coupled chaotic oscillators~\cite{Ott2013} in which extreme events called ``dragon kings'' (DKs)~\cite{Sornette2012} arise as a result of attractor bubbling~\cite{Ott2002,Gauthier1996,Venkataramani1996}. The precursor of extreme events is known for this model, and we demonstrate the effective suppression of extreme events by predicting the precursor.

The paper is organized as follows. In Sec. \ref{sec2}, we present the theoretical background of our algorithm. Section \ref{sec3} demonstrates numerical evidence of the proposed techniques for a system of globally coupled FHN neurons and for a system of two unidirectionally coupled chaotic oscillators. The conclusions are presented in Sec. \ref{sec5}.

\section{\label{sec2} Theoretical background} 

\subsection{\label{sec2a} Formulation of the problem}

In this paper, we consider the models of extreme events  described by ordinary differential equations of the form
\begin{equation}\label{model}
    \dot{\mathbf{x}}(t)=\mathbf{F}(\mathbf{x}(t),\mathbf{p})
\end{equation}
where $\mathbf{x}(t)$ is an $n$-dimensional state vector, $\mathbf{F}(\mathbf{x}(t),\mathbf{p})$ defines the vector field of the system, and $\mathbf{p}$ is a $n_p$-dimensional vector  of control parameters available for an external adjustment. We assume that in the absence of control this vector is zero, $\mathbf{p}=\mathbf{0}$. We also imagine that the output of the system is determined by a $d$-dimensional ($d\leq n$) vector observable   
\begin{equation}\label{output}
    \mathbf{s}(t)= \mathbf{h}(\mathbf{x}(t))
\end{equation}
that is a function  of the system state. Our goal is to predict extreme events, by observing the dynamics of the output variable $ \mathbf{s}(t)$, and to prevent them by applying small perturbations to the control parameter $\mathbf{p}$ at the instances of time preceding the extreme events.

\subsection{\label{sec2b} Prediction algorithm}

We assume that the model Eq.~\eqref{model} generating extreme events is unknown. We seek to predict extreme events or the precursors of extreme events (if  avalable) directly from the dynamics of the vector observable $\mathbf{s}(t)$. Within this limitation, the reservoir computer proposed in Ref.~\cite{jaeger04} is a promising tool for solving the problem. The ability of RC to provide model-free prediction has recently been demonstrated for various chaotic systems that do not exhibit extreme dynamics~\cite{Ott2017,Ott2017a,Ott2018,gaut2018,Ott2018b,Ott2018a}. 

\begin{figure}
\centering\includegraphics[width=0.45\textwidth]{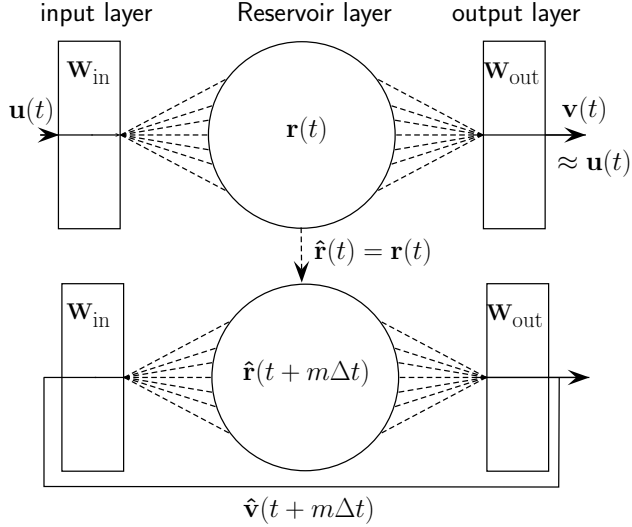}
\caption{\label{schema} Schematic diagram of the reservoir prediction setup. The top chart shows the listening reservoir based on an artificial neural network with $N$ neurons. The input vector $\mathbf{u}(t) \in \mathbb{R}^{d}$ is mapped to the reservoir state space $\mathbf{r}(t) \in \mathbb{R}^{N}$ by the input weight matrix $\mathbf{W}_{\textnormal{in}}$ [see Eqs.~\eqref{r_dyn} and \eqref{list_dyn}]. The reservoir output is $\mathbf{v}(t)=\mathbf{W}_{\textnormal{out}}^T \mathbf{r}(t)$, where   $\mathbf{W}_{\textnormal{out}}$ is the output matrix whose elements are obtained during the training stage to get an approximate equality  $\mathbf{v}(t)\approx \mathbf{u}(t)$. The bottom chart shows the predicting reservoir. Here the external input of the listening reservoir is replaced in a feedback loop by the post-processed reservoir output. After $M$ feedback iterations, the output of the predicting reservoir $\hat{\mathbf{v}}(t+\tau) \approx  \mathbf{u}(t+\tau)$ is ahead of the input signal by the time $\tau=M\Delta t$.
}
\end{figure}

The schematic diagram of the prediction algorithm utilized in this paper is shown in Fig~\ref{schema}. The algorithm uses two identical reservoir computers. Following Ref.~\cite{Ott2018}, we call them listening (top chart) and predicting (bottom chart) RCs. Each reservoir computer has three components: a linear input layer, a recurrent nonlinear reservoir network with $N$ dynamical reservoir nodes, and a linear output layer. The state of the listening reservoir is determined by the $N$-dimensional state vector $\mathbf{r}(t)$ that satisfies a discrete time deterministic model
\begin{equation}\label{r_dyn}
\mathbf{r}(t+\Delta t)=\mathbf{f}(\mathbf{r}(t), \mathbf{u}(t)),
\end{equation}
where $\Delta t$ is the step of time discretization, and $\mathbf{u}(t)$ is  the input signal of the reservoir. For the convenience of comparing the RC performances for different signals, we perform a linear transformation of
the output signal $\mathbf{s}(t)$ of the analyzed system \eqref{output} before applying it to the RC. Specifically, we preprocess the signal $\mathbf{s}(t)$ so that all the components of the input signal 
\begin{equation}\label{prepr}
u_{i}(t)=\left[s_{i}(t)-\langle s_{i}\rangle\right]/\sigma_{\textnormal{i}}, \quad i=1,\ldots,d
\end{equation}
that enter the listening reservoir have zero mean and unit variance. Here, $\langle s_{i}\rangle=\langle s_{i}(t)\rangle$ and $\sigma_{i}=\langle[s_{i}(t)-\langle s_{i}(t)\rangle]^2 \rangle^{1/2}$, where the angle brackets denote time average. There are many different ways to choose the nonlinear function $\mathbf{f}$ of the reservoir \cite{jaeger04,Ott2017,gaut2018,Appeltant2011,Haynes2015,Larger2017}. We select this function in the form used in the Refs.~\cite{jaeger04,Ott2017}:
\begin{equation}\label{list_dyn}
\mathbf{f}(\mathbf{r}, \mathbf{u})=(1-\alpha)\mathbf{r}+\alpha \tanh(\mathbf{A}\mathbf{r}+\mathbf{W}_{\textnormal{in}}\mathbf{u}+\xi \mathbf{1}).
\end{equation}
Here $\mathbf{A} \in \mathbb{R}^{N\times N}$ is the weighted adjacency matrix of the reservoir layer, and the input vector  $\mathbf{u}(t)$ is fed into the $N$ reservoir nodes using the linear input weight matrix $\mathbf{W}_{\textnormal{in}}\in \mathbb{R}^{N\times d}$. The parameter $0<\alpha\leq 1$ is a ``leakage'' rate that governs the evolution rate of the reservoir. Finally, $\xi \mathbf{1}$ is a bias term, where $\mathbf{1}$ denotes a vector of ones and $\xi$ is a scalar constant. 
For a vector $\mathbf{q}=[q_1,q_2,\ldots]^T$, the notation $\tanh(\mathbf{q})$ means the vector $[\tanh(q_1), \tanh(q_2), \ldots]^T$.

The matrices $\mathbf{A}$ and  $\mathbf{W}_{\textnormal{in}}$ are initially generated randomly and fixed. The parameters $D$, $\rho$, and $\sigma$ govern the random generation of these matrixes as follows. The adjacency matrix $\mathbf{A}$ is constructed  from a sparse random Erd\"{o}s-R\'{e}nyi matrix in which the average degree of a reservoir node is $D$, i.e., the non-zero elements form part of $D/N$. The values of non-zero elements are randomly generated from a uniform distribution in the interval $[-1, 1]$. Then the elements of $\mathbf{A}$ are rescaled so that the largest value of the magnitudes of its eigenvalues (the ``spectral radius'' of $\mathbf{A}$) becomes $\rho$. The elements of each column of the input matrix $\mathbf{W}_{\textnormal{in}}$ are randomly chosen from a uniform distribution in $[-\sigma, \sigma]$, where the parameter $\sigma$ represents the scalar input strength. The output of the listening reservoir is taken as a linear function of the reservoir state
\begin{equation}\label{v_out}
\mathbf{v}(t)=\mathbf{W}_{\textnormal{out}}^T\mathbf{r}(t),
\end{equation}
where $\mathbf{W}_{\textnormal{out}} \in \mathbb{R}^{N\times d} $ is the matrix of the output weights. This matrix is adjusted by the training process. Restricting the learning process by adjusting only the output weights (recall that the adjacency matrix and input vector are fixed) makes the reservoir computing approach advantageous compared to other artificial neural network approaches, since learning becomes computationally  feasible with relatively large $N$.

Our goal is to predict the signal $\mathbf{u}(t)$ corresponding to the output of the uncontrolled ($\mathbf{p}=\mathbf{0}$) system \eqref{model}.  We perform this task in two stages, which are called the stages of training and forecasting. At the training stage, we need only listening RC. Starting with the random initial state $\mathbf{r}(-t_{0})$, the reservoir evolves according to the Eq.~\eqref{r_dyn} with the input $\mathbf{u}(t)$ during a transient time $T_{\textnormal{tr}}$. The transient time is chosen large enough so that the state of the reservoir is essentially independent of its initial state by time $t=-t_0+T_{\textnormal{tr}} \equiv -T$. Then we record the $N_T=T/\Delta t$ reservoir states $\{\mathbf{r}(-T+\Delta t), \mathbf{r}(-T+2\Delta t), \ldots, \mathbf{r}(0) \}$  for the training interval $ -T<t\leq 0$. We train the network by choosing the output layer matrix $\mathbf{W}_{\textnormal{out}}$ so that the reservoir output $\mathbf{v}(t)$ approximates the input $\mathbf{u}(t)$ for $ -T<t\leq 0$. Specifically, we find the optimal output matrix $\mathbf{W}_{\textnormal{out}}$ by minimizing the following quadratic form:
\begin{equation}\label{minim}
 \sum_{k=1-N_T}^0 \left| \mathbf{W}_{\textnormal{out}}^T \mathbf{r}(k\Delta t)- \mathbf{u}(k\Delta t) \right|^2  + \beta \mathrm{Tr}(\mathbf{W}_{\textnormal{out}}^T\mathbf{W}_{\textnormal{out}}), 
\end{equation}
where $|\mathbf{q}|^2={\mathbf{q}}^T \mathbf{q}$ for $\mathbf{q}$ a vector. The second term in Eq.~\eqref{minim}, $\beta \mathrm{Tr}(\mathbf{W}_{\textnormal{out}}^T\mathbf{W}_{\textnormal{out}})$, is a regularization term introduced to avoid overfitting $\mathbf{W}_{\textnormal{out}}$, where $\beta$ is a small ``ridge regression parameter''~\cite{Yan2009,Ott2017,Ott2017a,Ott2018,gaut2018,Ott2018b,Ott2018a}. This modification of the ordinary linear least-squares regression is often called ridge regression or Tikhonov regularization. If the training is successful, the output $\mathbf{v}(t)=\mathbf{W}_{\textnormal{out}}^T\mathbf{r}(t)$ of the listening RC, with the $\mathbf{W}_{\textnormal{out}}$ determined in the training interval $ -T<t\leq 0$, should give a good approximation to the input, $\mathbf{v}(t) \approx \mathbf{u}(t)$, for $t>0$. 

To predict the input signal for times $t>0$, we use both RCs shown in Fig.~\ref{schema}. The state of the predicting RC (bottom chart), defined by the vector $\mathbf{\hat{r}}$, evolves autonomously with a feedback loop according to the equation
\begin{equation}\label{r_pred}
\mathbf{\hat{r}}(t+(m+1)\Delta t)=\mathbf{f}(\mathbf{\hat{r}}(t+m\Delta t), \hat{\mathbf{v}}(t+m\Delta t)),
\end{equation}
where
 \begin{equation}\label{v_out_pred}
    \hat{\mathbf{v}}(t+m\Delta t) =
    \begin{cases}
      \mathbf{u}(t) & \text{if $m=0$,} \\
      \mathbf{W}_{\textnormal{out}}^T\mathbf{\hat{r}}(t+m\Delta t) & \text{if $m=1,\ldots, M-1$.}
    \end{cases}
  \end{equation}
For the given time $t$, we perform $M$ iterations (changing $m$ from $0$ to $M-1$) of the Eq.~\eqref{r_pred}, using the initial condition $\mathbf{\hat{r}}(t)=\mathbf{r}(t)$. As a result, the output of the predicting RC allows us to estimate the predicted value of the input signal as
\begin{equation}\label{pred_sign}
\mathbf{u}(t+\tau) \approx \hat{\mathbf{v}}(t+\tau)\equiv\mathbf{W}_{\textnormal{out}}^T\mathbf{\hat{r}}(t+\tau),
\end{equation}
where
\begin{equation}\label{pred_time}
 \tau=M\Delta t
\end{equation}
is the prediction time. To estimate the quality of the prediction, we calculate the root mean square (RMS) error
\begin{equation}\label{RMS}
 R=[\langle\left[\hat{\mathbf{v}}(t+\tau)- \mathbf{u}(t+\tau)\right]^2\rangle]^{1/2}.
\end{equation}
Finally, the actual predicted output \eqref{output} of the system~\eqref{model} is given by 
\begin{equation}\label{pred_sign_orig}
s_{i}(t+\tau) \approx \hat{s}_{i}(t+\tau)\equiv \langle s_{i}\rangle+\sigma_{i} \hat{v}_{i}(t+\tau), \quad i=1,\ldots, d.
\end{equation}

 \subsection{\label{sec2c} Lyapunov exponents}

In the specific models presented below, extreme events manifest themselves  as rare large peaks in the dynamics of the scalar observable $s(t)$, which in general is a function of the output vector variable $\mathbf{s}(t)$ defined in Eq.~\eqref{output}, $s(t)=h_0(\mathbf{s}(t))=h_0(\mathbf{h}(\mathbf{x}(t)))\equiv h_1(\mathbf{x}(t))$. To assess the prediction horizon of extreme events, we will analyze the local stability of the uncontrolled ($\mathbf{p}=\mathbf{0}$) system \eqref{model} in close proximity to the large peaks. To do this, we turn to the concept of a local Lyapunov exponent (LE) \cite{abarb91,eck93}. We restrict ourselves by computing only the maximal local LE. This allows us to use  the simple Benettin algorithm \cite{benet80}. Our approach is as follows. We integrate jointly the uncontrolled system \eqref{model} and its linearized version:
\begin{subequations}\label{fx_df}
\begin{eqnarray}
    \dot{\mathbf{x}}(t)&=&\mathbf{F}(\mathbf{x}(t),0),\label{fx}\\
    \delta\dot{\mathbf{x}}(t)&=&\mathbf{J}(\mathbf{x}(t))\delta \mathbf{x}(t),\label{df}
\end{eqnarray}
\end{subequations}
where $\delta \mathbf{x}(t)$ is an infinitesimal deviation from the current system solution and  $\mathbf{J}(\mathbf{x}(t))$ is the Jacobian of the  system without control. First, we estimate the values of the observable $s(t_j)=h_1(\mathbf{x}(t_j))$ on the discrete time sequence $\{t_j=j \delta t\}_{j=1,\ldots, N_t}$ using a sufficiently small time step $\delta t$. Then we estimate the local maxima of the observable $s(t_j)$ and the large maxima that exceed some defined threshold value $s_{\textnormal{th}}$, we interpret as extreme events. The values of $t_j=t_{j_k}$, corresponding to these maxima, $\max_j [s(t_j)]>s_{\textnormal{th}}$, determine the moments of time $\{t_{j_k}\}_{k=1,\ldots,K_t<N_t}$ of extreme events. When solving linerized Eq. (\ref{df}), we rescale its variable $\delta\mathbf{x}(t_{j}) \rightarrow \delta\mathbf{x}(t_{j})/d_{j}$ at each time step $\delta t$, where $d_{j}=|\delta\mathbf{x}(t_{j})|/|\delta\mathbf{x}(t_{j-1})|$ determines the  relative growth of the deviation in the fastest direction in the phase space. To characterize the evolution of the deviation in the time interval $t_{j_k}-\tau_D<t<t_{j_k}-\tau_D+\tau_L$ near the $k$th extreme event occurring at time $t_{j_k}$, we define the local Lyapunov exponent as
\begin{equation}\label{Loc_Lyap}
\Lambda_{\text{loc}}^{(k)}(\tau_D,\tau_L)=\frac{1}{\tau_L}\sum_{j=j_k-D}^{j_k-N_D+L}\ln (d_j),
\end{equation}
where $N_D=\tau_D/\delta t$ is the number of time steps in the interval $[t_{j_k}-\tau_D, t_{j_k}]$ and $L=\tau_L/\delta t$ is the number of time steps in the interval $[t_{j_k}-\tau_D, t_{j_k}-\tau_D+\tau_L]$. The averaged local Lyapunov exponent over the extreme events is defined as
\begin{equation}\label{Loc_Lyap_aver}
\Lambda_\text{loc}(\tau_D,\tau_L)=\frac{1}{K_t}\sum_{k=1}^{K_t}\Lambda_\text{loc}^{(k)}(\tau_D,\tau_L).
\end{equation}
This exponent characterizes the maximum growth of deviations for a finite time interval, the beginning of which is placed at the time point  $\tau_D$ before the extreme event, and the length of the interval is $\tau_L$. Specifically, the relative growth of deviations in the given interval is estimated as $\exp[\Lambda_{\textnormal{loc}}(\tau_D,\tau_L) \tau_L]$.    

To compare the properties of the system in the neighborhood of extreme events with the global properties of the strange atractor, we also estimate the maximal global Lyapunov exponent $\Lambda$. We do this in a standard way, using the expression:
\begin{equation}\label{Max_Lyap}
\Lambda=\frac{1}{N_t\delta t}\sum_{j=1}^{N_t}\ln (d_j).
\end{equation}

In addition, we estimate the conditional Lyapunov exponents \cite{pecora1991,pyragas1997}  of the listening reservoir. They are determined from the linearized Eq.~\eqref{r_dyn}:
\begin{equation}\label{r_dyn_lin}
\delta\mathbf{r}(t+\Delta t)=D_1\mathbf{f}(\mathbf{r}(t), u(t))\delta \mathbf{r}(t).
\end{equation}
Here $\delta \mathbf{r}(t)$ is a small deviation from the solution $\mathbf{r}(t)$ of the listening reservoir driven with the input signal $u(t)$, and $D_1$ denotes the vector derivation of the function $\mathbf{f}$ with respect to its first argument. The conditional LEs (CLEs) are used to establish generalized synchronization (GS) \cite{pecora1990,rulkov1995,pyragas1998} of chaos in one-way coupled nonlinear dynamical systems. In the context of the present paper, the GS between the system \eqref{model} and the listening reservoir \eqref{r_dyn} means that the reservoir state $\mathbf{r}(t)$ becomes asymptotically a continuous function $\boldsymbol{\phi}$ of the system state $\mathbf{x}(t)$, i.e., $\mathbf{r}(t)\approx \boldsymbol{\phi}(\mathbf{x}(t))$ as $t\to\infty$. GS occurs if all conditional LEs are negative. As stated in Ref.~\cite{Ott2018}, GS is a necessary condition for the reservoir to predict the input signal $u(t)$.  A similar requirement for the reservoir, based on the so-called ``echo state property,'' was introduced by Jaeger \cite{jaeger2001}.

Here we use CLEs not only to establish the fact of generalized synchronization between the input signal and the reservoir, but also to estimate the characteristic time of the synchronization process. In our case the synchronization is disturbed by control perturbations. To ensure the correct prediction of extreme events in the presence of control perturbations, the characteristic time of the GS (determined by the maximal CLE) should be less than the characteristic time between extreme events. For the examples below, this condition is met and we can correctly predict and prevent all extreme events, despite the fact that the generalized synchronization is broken after each control action.

\section{\label{sec3} Numerical results}

In this section, we present the numerical results of forecasting and preventing extreme events in two different models. The first is a network of globally coupled FitzHugh-Nagumo neurons~\cite{Feudel2013,Feudel2014}, and the second is a system of two unidirectonally coupled chaotic oscillators~\cite{Ott2013}. In the first model, the precursors of extreme events are unknown, and we use a reservoir computer approach to predict extreme events directly from the scalar observable. In the second model, the precursor of extreme events is available, and instead of directly predicting extreme events we predict their precursors.

\subsection{\label{sec3a} FitzHugh-Nagumo systems}

First we consider extreme events in a network of $n$ diffusively coupled FitzHugh-Nagumo units described by the following differential equations~\cite{Feudel2014}:
%
\begin{eqnarray}
\dot{x}_{i} &=& x_{i}(a-x_{i})(x_{i}-1) - y_{i} + \frac{k}{n-1}\sum^{n}_{j=1}(x_{j}-x_{i})+p_i, \nonumber\\
\dot{y}_{i} &=& b_{i}x_{i}- c y_{i}, \quad i=1,\ldots,n.
\end{eqnarray}
%
Here, $a$, $b_i$, and $c$ are internal parameters of the units and
$k$ denotes the coupling strength. The parameters $a$ and $c$ are identical for all units, while $b$ is mismatched. Unlike to Ref.~\cite{Feudel2014}, here we introduce the control parameters $p_i$, which we use to prevent the extreme events. 

Similar to Ref.~\cite{Feudel2014}, we consider the following two sets of the parameters: 

(A) A small system of $n = 2$ units with $a=-0.025794$, $c=0.02$, $b_{1}=0.0135$, $b_{2}=0.0065$, and $k=0.129$.

(B) A large system of $n = 101$ units with $a=-0.02651$, $c=0.02$, $b_{i}=0.006+0.008(i-1)/(n-1)$ for $i=1,\ldots, n$, and $k=0.129$.

Below we study system A and system B separately. System A is discussed in details, and system B is used only to demonstrate  the applicability of our prediction and prevention control algorithm to large systems.

\subsubsection{System A}

\begin{figure}
\centering\includegraphics[width=0.45\textwidth]{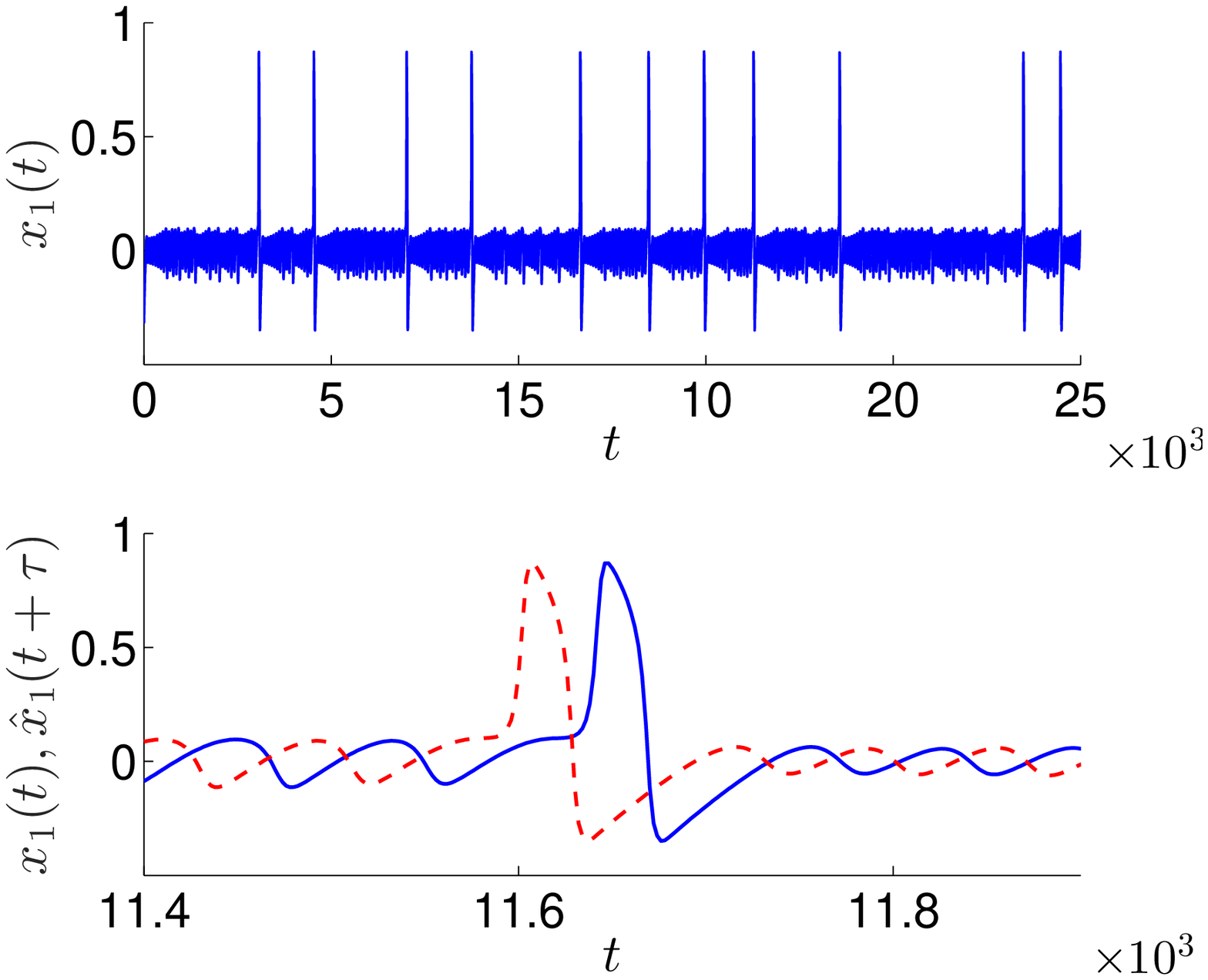}
\caption{\label{fhn1}  The output signal $s(t)=x_1(t)$ of the uncontrolled FHN system A (top) in a large time interval,  and (bottom) an episode of the same signal in a narrow time interval, showing one extreme event (solid blue curve). The dashed red curve shows the predicted signal $\hat{x}_1(t+\tau)$ obtained from the reservoir computer. The prediction time is $\tau=40$. The parameters of the reservoir computer are presented in Tab.~\ref{tab:table1}, second column.}
\end{figure}

For system A, consisting of only two FHN units, we choose the membrane potential of the first neuron, $x_1(t)$, as a scalar observable of the system, $s(t)=x_1(t)$. In Fig.~\ref{fhn1}, we show typical time series of this observable. Predominantly,  $x_1(t)$ exhibits low-amplitude oscillations, however, sometimes we observe large amplitude spikes that are at least six times higher than the amplitudes of the low-amplitude oscillations. These spikes can be qualified as extreme events since the observable exhibits unusually large values; the events are rare in comparison to the time scales of low-amplitude oscillations, and they are recurring.

\begin{figure}
\centering\includegraphics[width=0.45\textwidth]{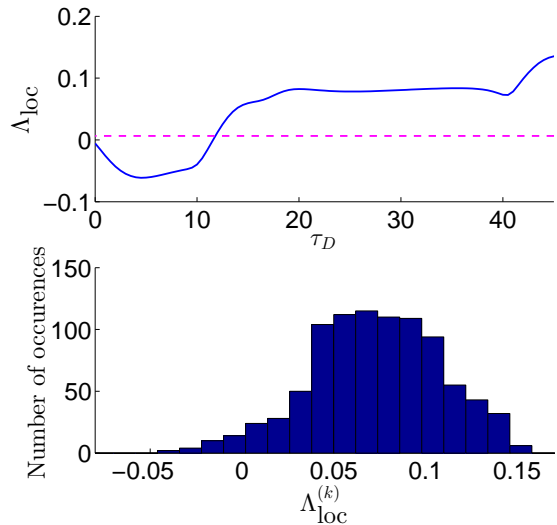}
\caption{\label{fhn_LE_stat} (Top) The averaged local LE $\Lambda_{\textnormal{loc}}(\tau_D,\tau_L)$ as a function of the time $\tau_D$  at fixed $\tau_L=40$ for the FHN system A. The horizontal dashed line shows the value of the maximal global LE $\Lambda \approx 0.0065$ estimated from Eq.~\eqref{Max_Lyap}. (Bottom) The histogram of the local LEs [Eq.~\eqref{Loc_Lyap}] at fixed $\tau_D=40$. The total number of extreme events is  $K_{t}=917$.}
\end{figure}

To estimate the growth of disturbances in the vicinity of extreme events, the local Lyapunov exponents were calculated as described in the Sec.~\ref{sec2c}.  Figure~\ref{fhn_LE_stat} shows the dependence of the averaged local LE $\Lambda_{\textnormal{loc}}(\tau_D,\tau_L)$  on the time $\tau_D$  for fixed $\tau_L=40$ (upper graph) and the histogram of the local LEs [Eq.~\eqref{Loc_Lyap}] for fixed $\tau_D=40$ (lower graph). We see that for $\tau_D >15$  the averaged local LE is much larger than the maximal global LE, whose value $\Lambda \approx 0.0065$ is shown in Fig.~\ref{fhn_LE_stat} with a horizontal dashed line. In the interval $20<\tau_D<40$, the averaged local LE weakly depends on $\tau_D$.  Here, its value $\Lambda_{\textnormal{loc}} \approx 0.073$ is almost ten times the global LE, and the relative growth of deviations close to the extreme events can be estimated as  $\exp(\Lambda_{\textnormal{loc}}\tau_{L})\approx 18.5$. For $\tau_D > 40$, the local LE increases even more, and therefore forecasting extreme events for times $\tau > 40$ should be very difficult.

We now discuss the results of predicting extreme events in the dynamics of the scalar observable $s(t)=x_1(t)$ of uncontrolled ($p_{1,2} = 0$) system. We used the algorithm described in the Sec~\ref{sec2b}, with a reservoir consisting of $N=1000$ neurons and other parameters presented in Tab.~\ref{tab:table1} (second column). We were able to correctly predict all extreme events with a time forecast $\tau = 40$, approximately equal to the width of the spikes of extreme events. In Fig.~\ref{fhn1} (lower graph) we show the episode of the prediction of one extreme event (dashed red curve), taken from the time series shown in the upper graph of the same figure.
\begin{table}
\label{table1}
  \begin{center}
    \caption{The parameters of the reservoir computer for three different systems considered in the paper: the second and third columns correspond to the FHN systems A and B, respectively, and the fourth column represents unidirectionally coupled chaotic systems (dragon kings).}
    \label{tab:table1}
    \begin{tabular}{c|c|c|c} 
      \textbf{Par.} & \textbf{FHN A} & \textbf{FHN B} & \textbf{DK}\\      
      \hline
      $N$ & 1000 & 2000 & 2000\\
      $D$ & 400 & 800 & 800\\
      $\rho$ & 1.6 & 1.6 & 1.2\\
      $\sigma$ & 1 & 1 & 2.8\\
      $\xi$ & 1 & 1 & 0.3\\
      $\alpha$ & 0.35 & 0.35 & 0.45\\
      $\beta$ & $10^{-8}$ & $10^{-4}$ & $10^{-7}$\\
      $\Delta t$ & $2$ & $2$ & $2$\\ 
      $M$ & $20$ & $8$ & $4$\\ 
      $T_{\textnormal{tr}}$ & $1000$ & $2000$ & $1000$\\
      $T$ & $8000$ & $36000$ & $60000$\\     
     \end{tabular}
     \end{center}
\end{table}

\begin{figure}
\centering\includegraphics[width=0.45\textwidth]{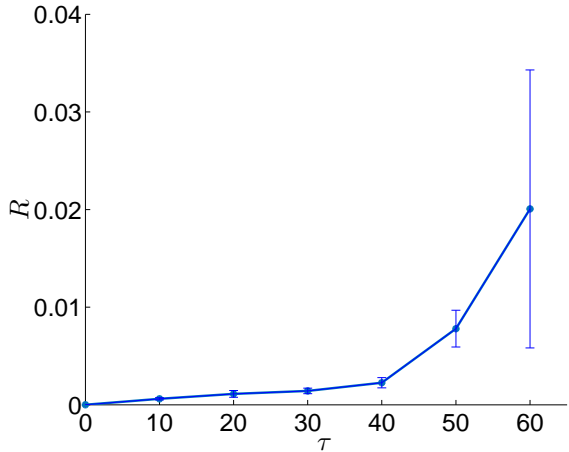}
\caption{\label{D_fhn} The RMS error $R$ [Eq.~\eqref{RMS}] of the the forecast versus the forecast time $\tau$ for the FHN system A. 
}
\end{figure}

Figure~\ref{D_fhn} demonstrates the forecasting horizon of extreme events. Here we depict the RMS error $R$ [Eq.~\eqref{RMS}]  of the forecast  depending on  the forecast  time $\tau$. Each plotted point in this figure is the median of $20$ trials using the same output signal, with each trial using an independent random realization of the reservoir system, and the error bars indicate the range in the performance of the RC. We see that the RMS error is extremely small for $\tau \le 40$ and  increases rapidly for $\tau$ over $40$. This is consistent with local LE analysis; the rapid growth of local LE in Fig.~\ref{fhn_LE_stat} also starts when $\tau_D>40$.

\begin{figure}
\centering\includegraphics[width=0.45\textwidth]{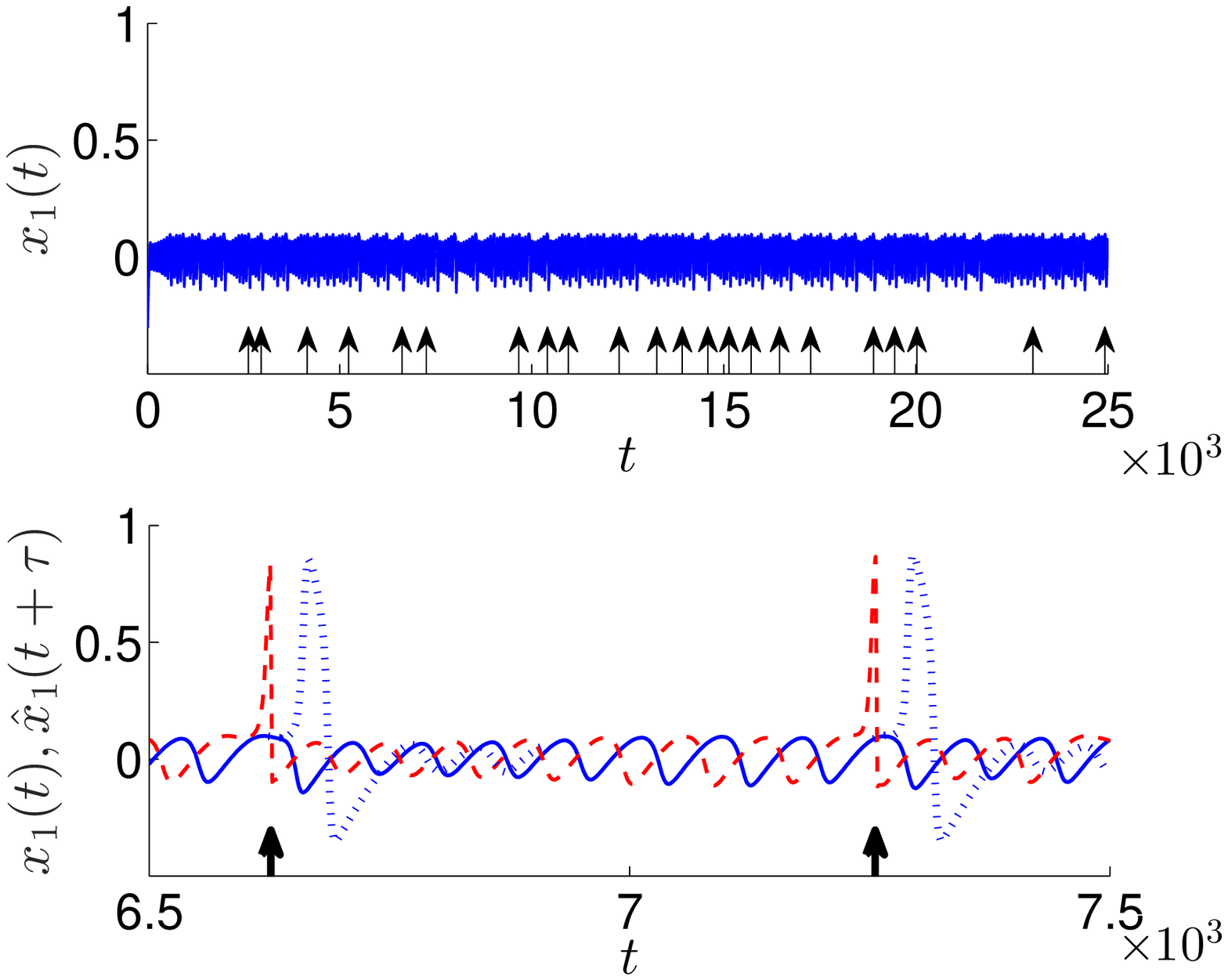}
\caption{\label{fhn2t}  Suppression of extreme events in the FHN system A. (Top) Dynamics of the output variable $s(t)=x_1(t)$ in the presence of control. The initial conditions of the FHN system are the same as in Fig.~\ref{fhn1}. The arrows indicate the moments of supply of control pulses. The amplitude of the pulses is $\varepsilon =10^{-3}$. (Bottom) The extended time interval extracted from the upper graph showing the details of suppression of two extreme events. The solid blue curve shows the actual output signal $x_1(t)$ under control. Blue dotted curves represent the solutions of the FHN system, near extreme events that would occur without control. The red dashed curve shows the predicted signal $\hat{x}_1(t+\tau)$ obtained from the RC system in the presence of control.}
\end{figure}

Using the predicted signal $\hat{s}(t+\tau)\equiv\hat{x}_1(t+\tau) \approx x_1(t+\tau)$ obtained from RC, we were able to suppress extreme events with small external perturbations. We activated these perturbations at times preceding extreme events. Specifically, our algorithm for suppressing extreme events is as follows. We identify the moment $t_k^{*}$ preceding the $k$th extreme event as the moment when the predicted signal crosses from below to above  a threshold value $s^*=0.7$, i.e., when $\hat{x}_1(t_k^{*}+\tau)>s^*$ and $\hat{x}_1(t_k^{*}+\tau-\Delta t)<s^*$. At every moment $t_k^{*}$, we perturb the first neuron with a small negative delta pulse, $p_1(t)=-\varepsilon\sum_k \delta(t-t_k^*)$, while the second neuron remains unperturbed, $p_2(t)\equiv 0$. The results of such a prediction and prevention algorithm with a perturbation amplitude of $\varepsilon = 10^{-3}$ are presented in Fig.~\ref{fhn2t}. The initial conditions of the FHN system are the same as in the Fig. \ref{fhn1}. Now the upper graph shows the actual dynamics under control. The arrows indicate the moments $t_k^*$ at which the control perturbation was applied. We see that all extreme events are totally suppressed and the system displays only low amplitude oscillations. The lower graph shows the extended time interval of the upper graph. Here we see the details of the suppression of two extreme events. The solid blue curve shows the actual output signal $x_1(t)$ under control and blue dotted curves show the solutions of the FHN system, near extreme events that would occur without control. The red dashed curve shows the predicted signal $\hat{x}_1(t+\tau)$ obtained using the RC system in the presence of control. 

Finally, we discuss the problem of generalized synchronization between the FHN system and the listening reservoir. As indicated in the Sec.~\ref{sec2c}, the presence of GS is a prerequisite for the functioning of the reservoir computer. Our numerical analysis of conditional Lyapunov exponents for the uncontrolled FHN system showed that all CLEs are negative with the maximal CLE equal to $\Lambda_c \approx - 0.036$. Thus, the necessary condition for the GS between the uncontrolled FHN system and the reservoir is fulfilled. In the presence of control, we need to fulfill an additional requirement. Since each control pulse violates the GS, the reservoir computer cannot correctly predict the input signal immediately after the control pulse. However, if extreme events are rare and small, then GS can recover between control pulses, and we can get the correct forecast of extreme events even with control. The  characteristic time of the GS can be estimated as the reciprocal of the maximal CLE, $T_{gs} = 1/|\Lambda_{c}|\approx 27.78$. In our case, this time is significantly less than the average time distance $T_{ee} \approx 100$ between two adjacent extreme events. Thus, the reservoir is fast enough  to restore its adequate performance between control perturbations.  

\subsubsection{System B}
\begin{figure}
\centering\includegraphics[width=0.45\textwidth]{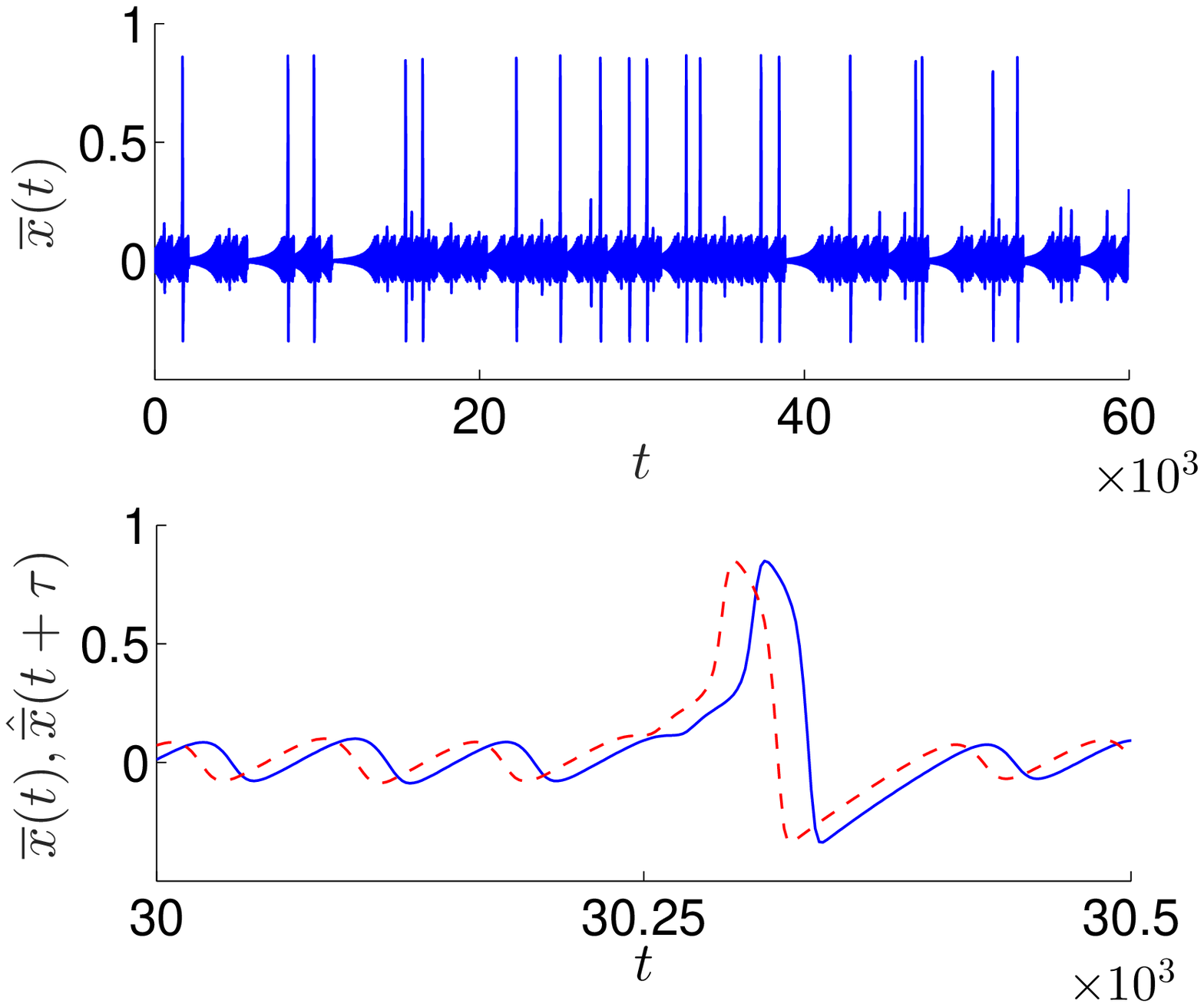}
\caption{\label{fig_B2} The output signal $s(t)=\bar{x}(t)$ of the uncontrolled FHN system B (top) in a large time interval  and (bottom) an episode of the same signal in a narrow time interval, showing one extreme event (solid blue curve). The dashed red curve shows the predicted signal $\hat{\bar{x}}(t+\tau)$ obtained from the reservoir computer. The prediction time is $\tau=16$. The parameters of the reservoir computer are presented in Tab.~\ref{tab:table1}, third column. 
}
\end{figure}

Now we show that our prediction and prevention algorithm also works for the large FHN system B consisting of $n = 101$ units. In this case, we take the mean-field potential as a scalar observable of the system:
\begin{equation}\label{FHN_out}
s(t)= \bar{x}(t) \equiv \frac{1}{n}\sum^{n}_{j=1}x_{j}(t)
\end{equation}
In the absence of control, a typical dynamics of this observable is shown in Fig.~\ref{fig_B2} (upper graph).  Using the reservoir computer with the parameters presented in the Tab.~\ref{tab:table1} (third column), we were able to correctly predict all extreme events with a time forecast $\tau = 16$, approximately equal to half the width of the spikes of extreme events. The episode of predicting one extreme event is shown in the lower graph (the dashed red curve). 

\begin{figure}
\centering\includegraphics[width=0.45\textwidth]{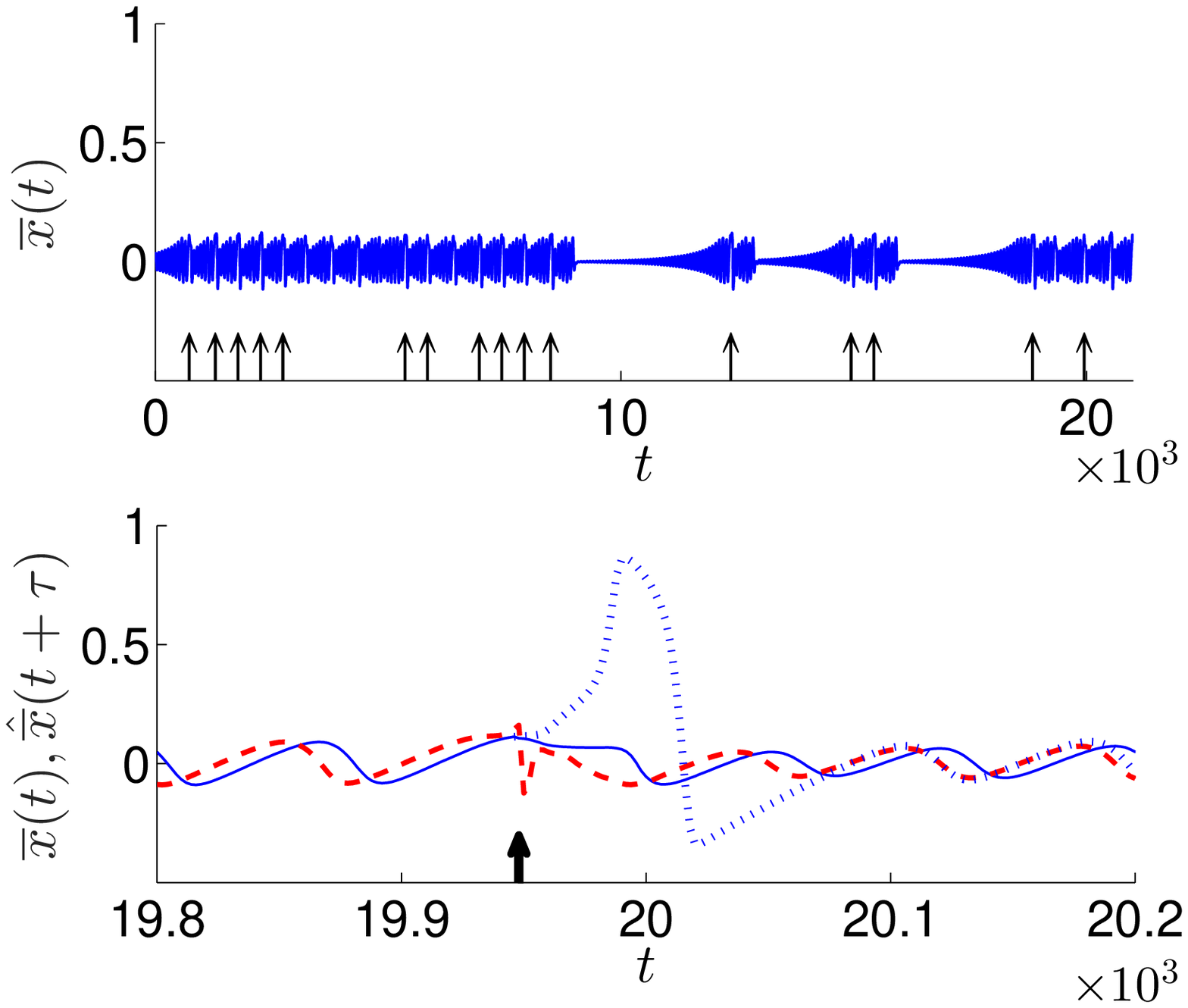}
\caption{\label{fig_B1}  Suppression of extreme events in the FHN system B. (Top) Dynamics of the output variable $s(t)=\bar{x}(t)$ in the presence of control with the initial conditions the same as in Fig.~\ref{fig_B2}. The arrows indicate the moments of supply of control pulses. The amplitude of the pulses is $\varepsilon =5\cdot 10^{-3}$. (Bottom) The extended time interval of the upper graph showing the details of suppression of one extreme event. The marking of the curves is the same as in Fig.~\ref{fhn2t}.
}
\end{figure}

Although the forecast time of extreme events for system B is shorter than for system A, it is large enough to suppress extreme events with small perturbations. The results of suppressing extreme events in system B are shown in Fig.~\ref{fig_B1}. Here we used a similar algorithm as for system A. The moments $t_k^*$ preceding extreme events were identified from the predicted signal $\hat{s}(t+\tau)\equiv\hat{\bar{x}}(t+\tau)$ obtained from RC by using the threshold $s^*=0.14$. Now the controlling delta pulses  were applied at the
moments $t_k^*$ to all neurons, i.e., we used  the control perturbation of the form: $p_j(t)=-\varepsilon\sum_k \delta(t-t_k^*)$ for $j=1,\ldots, n$. After each control event, the system was not perturbed for the next $25\Delta t$ time units. With a sufficiently small perturbation amplitude $\varepsilon = 5 \cdot 10^{-3}$, we were able to suppress all extreme events, as can be seen from the upper graph. Details of suppressing one extreme event are shown in the lower graph. 

\subsection{\label{sec4} Unidirectionally coupled chaotic systems}

As a final example, we consider the model of extreme events described by two nearly identical unidirectionally coupled chaotic oscillators in a master (mnemonic M) and slave (S) configuration~\cite{Ott2013}:
\begin{subequations}\label{ms_dyn}
\begin{eqnarray}
\dot{\mathbf{x}}_{M} &=& \mathbf{F}_{M}(\mathbf{x}_{M}),  \label{ms_dyna}\\
\dot{\mathbf{x}}_{S} &=& \mathbf{F}_{S}(\mathbf{x}_{S})+c\mathbf{K}(\mathbf{x}_{M}-\mathbf{x}_{S})+\mathbf{p}.\label{ms_dynb}
\end{eqnarray}
\end{subequations}
Here $\mathbf{x}_{M}$ and $\mathbf{x}_{S}$ are three-dimensional (3D) state vectors of the master and slave subsystems, respectively, while $\mathbf{F}_{M}(\mathbf{x}_{M})$ and $\mathbf{F}_{S}(\mathbf{x}_{S})$ are the corresponding flows. The flows are described by the same vector function, but with slightly mismatched parameters, so $\mathbf{F}_{M}(\mathbf{x}) \approx \mathbf{F}_{S}(\mathbf{x})$. The term $c\mathbf{K}(\mathbf{x}_{M}-\mathbf{x}_{S})$ is responsible for the interaction between the subsystems, where $c$ determines the strength of the interaction and  $\mathbf{K}$ is the coupling matrix. Finally, $\mathbf{p}$ is a 3D vector of control parameters. 

First we discuss the solutions of the system without control, when $\mathbf{p}=\mathbf{0}$. Assume that the master and slave subsystems are identical,  $\mathbf{F}_{M}(\mathbf{x}) = \mathbf{F}_{S}(\mathbf{x})$. Then  in a six-dimensional (6D) phase space spanned by $(\mathbf{x}_M,\mathbf{x}_S$) there exists a 3D invariant (synchronization) manifold $\mathbf{x}_S=\mathbf{x}_M$. For appropriate values of $c$ and $\mathbf{K}$, the coupled oscillators synchronize their behavior and the trajectory settles on this synchronization manifold. However, small noise or slight parameter mismatch in such systems can lead to attractor bubbling~\cite{Ott2002,Gauthier1996,Venkataramani1996}, which manifests itself in the dynamics of the state vector $\mathbf{x}_{\perp}=(\mathbf{x}_{M}-\mathbf{x}_{S})/2$ transverse to the synchronization manifold. Long intervals of high-quality synchronization, where $|\mathbf{x}_{\perp}(t)| \approx 0$ are interrupted at irregular times by large, brief desynchronization events, where $|\mathbf{x}_{\perp}(t)|$ shows high-amplitude spikes, which can be interpreted as extreme events. In Ref.~\cite{Ott2013}, these events were clasified as dragon kings~\cite{Sornette2012}, since the event-size distribution deviates significantly upward from a power law in the tail. 

\begin{figure}
\centering\includegraphics[width=0.45\textwidth]{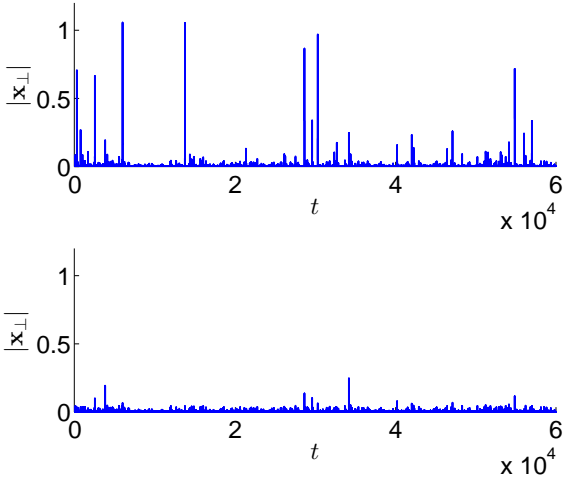}
\caption{\label{ott1} Suppression of extreme events in unidirectionally coupled chaotic systems. (Top) Extreme events in the dynamics of a transverse vector variable $|\mathbf{x}_{\perp}|$ without control.  
(Bottom) Dynamics of the variable $|\mathbf{x}_{\perp}|$ in the presence of control defined by the Eq.~\eqref{contr_our} with parameter values   $c_{DK}=0.55$, $|\mathbf{x}_{M}|_{\textnormal{th}}=0.25$ and $\tau=6$.
}
\end{figure}

A specific example of the system~\eqref{ms_dyn} was implemented in Ref.~\cite{Ott2013} using electronic circuits for which the state variables of the master and slave systems are respectively $\mathbf{x}_{M}=[V_{1M}, V_{2M}, I_{M}]^{T}$ and $\mathbf{x}_{S}=[V_{1S}, V_{2S}, I_{S}]^{T}$, and the corresponding flows $\mathbf{F}_{j}(\mathbf{x}_{j})$ for $j=M,S$ take the form 
\begin{equation}\label{ms_syst}
\mathbf{F}_{j} = 
\left[ \begin{array}{l}
V_{1j}/R_{1j}-g(V_{1j}-V_{2j}) \\
g(V_{1j}-V_{2j})-I_{j} \\
V_{2j}-R_{4j}I_{j}
\end{array} \right],
\end{equation}
where
\begin{equation}\label{ms_g}
g(V)=V/R_{2j}+I_{rj}[\exp(\alpha_{fj}V)-\exp(-\alpha_{rj}V)].
\end{equation}
Below we present the numerical results for the following values of the system parameters. For both subsystems $(j=M,S)$, we take the same parameters
$R_{1j}=1.2$, $R_{2j}=3.44$, $I_{rj}=22.5\cdot 10^{-6}$, $\alpha_{fj}=11.6$, and $\alpha_{rj}=11.6$, but we assume that the parameters $R_{4j}$  are slightly different:  $R_{4M}=0.193$ and $R_{4S}=0.194$. The coupling between the oscillators is carried out through the $3 \times 3$  matrix $\mathbf{K}$, all elements of which are equal to zeros, except  $K_{22} = 1$, and the coupling strength is taken to be $c = 4.4$. For given parameter values, the typical temporal evolution of the variable  $|\mathbf{x}_{\perp}|$ of the system~\eqref{ms_dyn} without control is shown in Fig.~\ref{ott1} (upper graph). Bubbling is indicated by long excursions of high-quality synchronization 
($|\mathbf{x}_{\perp}|$ is proportional to the difference of the mismatched parameters $R_{4j}$) interspersed by brief
desynchronization events where $|\mathbf{x}_{\perp}|$ takes on a large value -- an extreme event.

For the coupled oscillators considered here, it was found that a saddle-type fixed point $\mathbf{x}_S=\mathbf{x}_M = \mathbf{0}$ lying on the synchronization manifold is exceedingly transversely unstable and is the main source of the largest bubbles (the dragon kings)~\cite{Ott2013}. The influence of a fixed point in the dynamics makes it possible to predict the occurrence of a large event by observing in real time $\mathbf{x}_M$, which is equal to $\mathbf{x}_S$ when the subsystems are synchronized. A drop of the variable $|\mathbf{x}_M|$ of the master system below an empirically determined threshold value $|\mathbf{x}_{M}|_{\textnormal{th}}$ can serve as a precursor of an extreme event appearing in the dynamics of the variable  $|\mathbf{x}_{\perp}|$. A smaller threshold is associated with a larger event size. Using this precursor, the authors of Ref.~\cite{Ott2013} designed a feedback method to suppress DKs. The algorithm is based on occasional proportional feedback of small perturbations to the slave oscillator when  $|\mathbf{x}_M(t)|<|\mathbf{x}_{M}|_{\textnormal{th}}$. Specifically, they assumed that only the first component of the control vector is nonzero, $\mathbf{p}=[p_1, 0, 0]^T$, and used the following control law:
\begin{equation}\label{contr_Ott}
p_1(t) = c_{DK}\Theta[|\mathbf{x}_{M}|_{\textnormal{th}}-|\mathbf{x}_{M}(t)|][V_{1M}(t)-V_{1S}(t)], 
\end{equation}
where $c_{DK}$ is the strength of the control perturbation and $\Theta[|\mathbf{x}_{M}|_{\textnormal{th}}-|\mathbf{x}_{M}(t)|]$ --- Heaviside step function equal to one for $|\mathbf{x}_M(t)|<|\mathbf{x}_{M}|_{\textnormal{th}}$ and otherwise equal to zero. This function determines the time intervals in which the control perturbation is activated. 

\begin{figure}
\centering\includegraphics[width=0.45\textwidth]{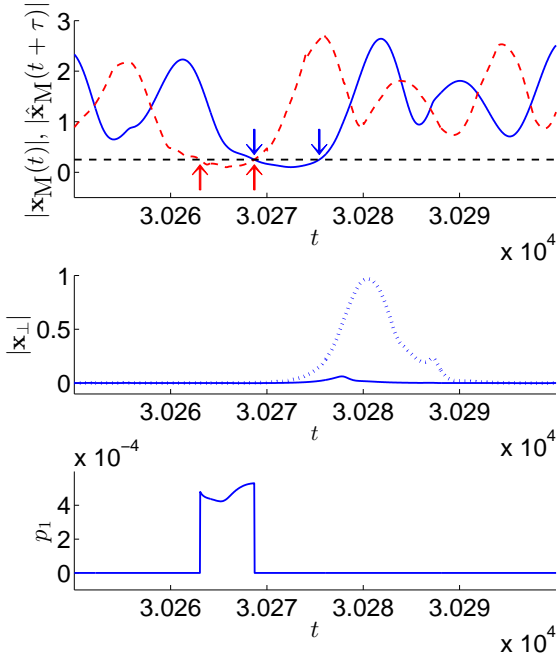}
\caption{\label{ott2} Details of the suppression of one dragon king. (Top) Dynamics of the  variable $|\mathbf{x}_{M}(t)|$ (blue solid curve) and its predicted value $|\mathbf{\hat{x}}_{M}(t+\tau)|$ (red dashed curve) obtained from the reservoir computer. Horizontal dashed line shows the threshold $|\mathbf{x}_{M}|_{\textnormal{th}}=0.25$. The blue down arrows and red up arrows indicate time intervals of activating the control perturbation \eqref{contr_Ott} and \eqref{contr_our}, respectively. (Middle) Dynamics of the variable  $|\mathbf{x}_{\perp}|$ without control (dotted curve) and in the presence of control (solid curve) defined by the Eq.~\eqref{contr_our}. (Bottom) Dynamics of control perturbation~\eqref{contr_our}.}
\end{figure}

Here we show that the control algorithm~\eqref{contr_Ott} can be improved by implementing the predicted values of the precursor obtained using the reservoir computer. Our modification of the control law~\eqref{contr_Ott} is as follows:
\begin{equation}\label{contr_our}
p_1(t) = c_{DK}\Theta[|\mathbf{x}_{M}|_{\textnormal{th}}-|\mathbf{\hat{x}}_{M}(t+\tau)|][V_{1M}(t)-V_{1S}(t)]. 
\end{equation}
In the Heaviside function argument, we replace the current output variable $|\mathbf{x}_{M}(t)|$ of the master system with its predicted value $|\mathbf{\hat{x}}_{M}(t+\tau)|$ obtained from the reservoir computer. Unlike the previous examples, where only scalar signals were predicted, here we predict the three-dimensional vector variable $\mathbf{x}_{M}(t)$. Another difference is that here we use a slightly modified version of the reservoir computer,  recently proposed in the Ref.~\cite{Ott2018b}. The modification concerns the output of the reservoir computer. In the Eq.~\eqref{v_out} that defines the output signal, we replace the reservoir state vector $\mathbf{r}=[r_1,\ldots, r_N]^T$ with the extended $2N$ dimensional vector $\mathbf{r}_{\textnormal{ex}}=[r_1,\ldots, r_N, r_1^2,\ldots, r_N^2]^T$. Then the dimension of the output matrix $\mathbf{W}_{\textnormal{out}}$ becomes $2N\times d$, and the Eq.~\eqref{v_out} reads $\mathbf{v}(t)=\mathbf{W}_{\textnormal{out}}^T\mathbf{r}_{\textnormal{ex}}(t)$. With this modification, all operations described in Sec.~\ref{sec2b} remain the same. The values of the reservoir parameters that we used in our simulations are presented in Tab.~\ref{tab:table1}, fourth column. We predict the behavior of the master system for $M=4$ discrete time steps $\Delta t=2$. However, when searching for the intersection points of the predicted signal with the threshold $|\mathbf{x}_{M}|_{\textnormal{th}}$ we need to interpolate the signal between discrete  points. This leads to the loss of one time step in the prediction, so the prediction time in a real experiment with the control law~\eqref{contr_our} is $\tau=(M-1)\Delta t=6$. The bottom graph in Fig.~\ref{ott1} shows the dynamics of the variable $|\mathbf{x}_{\perp}|$ under the control algorithm~\eqref{contr_our} with strength $c_{DK}=0.55$ and threshold $|\mathbf{x}_{M}|_{\textnormal{th}}=0.25$. We see that all large extreme events that occur in the dynamics of an uncontrolled system (upper graph) are successfully suppressed. Several low-amplitude bubbles remained due to the finite forecast error of the reservoir computer.
Suppression is achieved by tiny perturbations --- the amplitude of the control parameter $p_1(t)$ is of the order of $5\times 10^{-4}$ (see the bottom graph in Fig.~\ref{ott2}).

Figure~\ref{ott2} shows the details of suppressing one DK extracted from Fig.~\ref{ott1}. The upper graph shows the dynamics of the  variable $|\mathbf{x}_{M}(t)|$ (blue solid curve) and its predicted value $|\mathbf{\hat{x}}_{M}(t+\tau)|$ (red dashed curve) obtained from the reservoir computer. The threshold $|\mathbf{x}_{M}|_{\textnormal{th}}$ is shown by the horizontal dashed line. The middle graph shows the dynamics of the variable  $|\mathbf{x}_{\perp}|$ without control (dotted curve) and in the presence of control (solid curve), while the lower graph shows the dynamics of the perturbation. The time interval in which the control is activated is determined by the master system --- it corresponds to the time interval in which the predicted signal $|\mathbf{\hat{x}}_{M}(t+\tau)|$ is below the threshold value $|\mathbf{x}_{M}|_{\textnormal{th}}$. In the upper graph, this interval is indicated with red up arrows. In the case of an ideal prediction, the length of this interval coincides with the length of the corresponding interval (marked with blue down arrows) of the control law~\eqref{contr_Ott}, but in our case the perturbation is activated for a time $\tau$ earlier than in the control law~\eqref{contr_Ott}.

Earlier activation of control requires less energy to suppress an upcoming extreme event. To compare the effectiveness of the control laws~\eqref{contr_Ott} and \eqref{contr_our}, we calculated for both of them the average control energy needed to suppress one extreme event:
\begin{equation}\label{av_pow}
E=\frac{1}{N_{E}}\int^{T_E}_{0} [p_1(t)]^2 dt,
\end{equation}
where $[0, T_E]$ is a sufficiently long time interval containing a sufficiently large number $N_E$ of control events. For the given parameter values, our algorithm~\eqref{contr_our} gives $E \approx 3.7\times 10^{-7}$, while the algorithm~\eqref{contr_Ott} for the same parameter values, gives $E \approx 4.3 \times 10^{-5}$. Thus, the proposed modification~\eqref{contr_our} allows us to reduce the average energy of suppression of extreme events by two orders of magnitude. 

As a final remark, we note the important advantage of the unidirectionally coupled systems considered here over the systems analyzed in the Sec.~\ref{sec3a}. Here forecasting is necessary only for the master system, and the control signal is supplied only to the slave system. Therefore, the control signal does not affect the dynamics of the master system and does not violate the GS between the RC and the master system. This makes predicting and preventing extreme events easier than in the Sec.~\ref{sec3a}.

\section{\label{sec5} Conclusions}

We have shown that a reservoir computer is an effective tool for predicting extreme events in chaotic systems with {\it a~priori} unknown models. Using this prediction, various control algorithms can be implemented to suppress unwanted extreme events by applying a small perturbation to the system before an extreme event occurs.
We have demonstrated such a prediction and prevention strategy for two chaotic systems exhibiting the dynamics of extreme events.

The first is a system of globally coupled FitzHugh-Nagumo neurons~\cite{Feudel2013,Feudel2014}. 
In the case of only two interconnected neurons, we presented a detailed analysis of the forecasting algorithm using various numerical criteria. 
Our analysis showed that the average local LE characterizing the divergence of trajectories near extreme events is much larger than the global LE of the strange attractor. Despite this, the reservoir computer provided a long-term forecast of extreme events, so we managed to suppress them by applying only tiny perturbations to the system. We also analyzed a  generalized synchronization between the FHN system and the listening reservoir, which is a prerequisite for the functioning of the reservoir computer. By calculating the conditional LE, we showed that the characteristic time of the GS is significantly less than the characteristic time between two adjacent extreme events. Although the GS is broken after each control action, it manages to recover between extreme events. We have demonstrated the effectiveness of the prediction and prevention algorithm not only for two interconnected neurons, but also for a large-scale system of globally coupled FHN neurons.

The second system, analyzed in this paper, is two almost identical unidirectionally connected chaotic oscillators~\cite{Ott2013} in which extreme events, known as dragon kings~\cite{Sornette2012}, arise as a result of attractor bubbling. The precursor of extreme events is known for this system, and we used this advantage in our prediction and prevention algorithm. Instead of directly predicting extreme events, here we turned to the reservoir computer to predict the precursor. This allowed us to reduce the average energy required to suppress the dragon kings by two orders of magnitude compared to the previously propsed control strategy~\cite{Ott2013} without predicting the precursor.

\section*{Acknowledgment}

This work is supported by grant  
No. P-MIP-20-11 of the Research Council of Lithuania.

\bibliographystyle{elsarticle-num}
\bibliography{references}

\begin{thebibliography}{10}
\expandafter\ifx\csname url\endcsname\relax
  \def\url#1{\texttt{#1}}\fi
\expandafter\ifx\csname urlprefix\endcsname\relax\def\urlprefix{URL }\fi
\expandafter\ifx\csname href\endcsname\relax
  \def\href#1#2{#2} \def\path#1{#1}\fi

\bibitem{Sergio2006}
S.~Albeverio, V.~Jentsch, H.~Kantz (Eds.), Extreme Events in Nature and
  Society, Springer, Berlin, 2006.

\bibitem{Disthe2008}
K.~Dysthe, H.~E. Krogstad, P.~M\"{u}ller, Oceanic rogue waves, Annual Review of
  Fluid Mechanics 40~(1) (2008) 287--310.

\bibitem{Ohnaka2013}
M.~Ohnaka, The Physics of Rock Failure and Earthquakes, Cambridge University
  Press, New York, 2013.

\bibitem{Crucitti2004}
P.~Crucitti, V.~Latora, M.~Marchiori, Model for cascading failures in complex
  networks, Phys. Rev. E 69 (2004) 045104.

\bibitem{Sornette2003}
D.~Sornette, Why Stock Markets Crash, Princeton University Press, Princeton,
  NJ, 2003.

\bibitem{Engel2007}
J.~J. Engel, T.~A. Pedley (Eds.), Epilepsy: A Comprehensive Textbook, 2nd ed.,
  Lippincott, Williams \& Wilkins, Philadelphia, PA, 2007.

\bibitem{Forgoston2018}
E.~Forgoston, R.~Moore, A primer on noise-induced transitions in applied
  dynamical systems, SIAM Review 60~(4) (2018) 969--1009.

\bibitem{Zaldivar2005}
J.-M. Zald\'{i}var, J.~Bosch, F.~Strozzi, J.~P. Zbilut, Early warning detection
  of runaway initiation using non-linear approaches, Communications in
  Nonlinear Science and Numerical Simulation 10~(3) (2005) 299--311.

\bibitem{Farazmand2016}
M.~Farazmand, T.~P. Sapsis, Dynamical indicators for the prediction of bursting
  phenomena in high-dimensional systems, Phys. Rev. E 94 (2016) 032212.

\bibitem{faraz18}
M.~Farazmand, T.~P. Sapsis, Extreme events: Mechanisms and prediction, Appl.
  Mech. Rev. 71~(5) (2019) 050801.

\bibitem{jaeger04}
H.~Jaeger, H.~H., Harnessing nonlinearity: Predicting chaotic systems and
  saving energy in wireless communication, Science 304 (2004) 78.

\bibitem{verst2007}
D.~Verstraeten, B.~Schrauwen, M.~D'Haene, D.~Stroobandt, An experimental
  unification of reservoir computing methods, Neural Networks 20~(3) (2007)
  391--403.

\bibitem{Mantas09}
M.~Luko\v{s}evi\v{c}ius, H.~Jaeger, Reservoir computing approaches to recurrent
  neural network training, Computer Science Review 3~(3) (2009) 127--149.

\bibitem{Wyffels2010}
F.~Wyffels, B.~Schrauwen, A comparative study of reservoir computing strategies
  for monthly time series prediction, Neurocomputing 73~(10) (2010) 1958--1964.

\bibitem{Luk2012}
M.~Luko{\v{s}}evi{\v{c}}ius, H.~Jaeger, B.~Schrauwen, Reservoir computing
  trends, KI - K{\"u}nstliche Intelligenz 26~(4) (2012) 365--371.

\bibitem{Ott2017}
Z.~Lu, J.~Pathak, B.~Hunt, M.~Girvan, R.~Brockett, E.~Ott, Reservoir observers:
  Model-free inference of unmeasured variables in chaotic systems, Chaos: An
  Interdisciplinary Journal of Nonlinear Science 27~(4) (2017) 041102.

\bibitem{Ott2017a}
J.~Pathak, Z.~Lu, B.~R. Hunt, M.~Girvan, E.~Ott, Using machine learning to
  replicate chaotic attractors and calculate lyapunov exponents from data,
  Chaos: An Interdisciplinary Journal of Nonlinear Science 27~(12) (2017)
  121102.

\bibitem{Ott2018}
Z.~Lu, B.~R. Hunt, E.~Ott, Attractor reconstruction by machine learning, Chaos:
  An Interdisciplinary Journal of Nonlinear Science 28~(6) (2018) 061104.

\bibitem{gaut2018}
D.~Canaday, A.~Griffith, D.~J. Gauthier, Rapid time series prediction with a
  hardware-based reservoir computer, Chaos: An Interdisciplinary Journal of
  Nonlinear Science 28~(12) (2018) 123119.

\bibitem{Ott2018b}
J.~Pathak, B.~Hunt, M.~Girvan, Z.~Lu, E.~Ott, Model-free prediction of large
  spatiotemporally chaotic systems from data: A reservoir computing approach,
  Phys. Rev. Lett. 120 (2018) 024102.

\bibitem{Ott2018a}
J.~Pathak, A.~Wikner, R.~Fussell, S.~Chandra, B.~R. Hunt, M.~Girvan, E.~Ott,
  Hybrid forecasting of chaotic processes: Using machine learning in
  conjunction with a knowledge-based model, Chaos: An Interdisciplinary Journal
  of Nonlinear Science 28~(4) (2018) 041101.

\bibitem{pyr2019}
T.~Pyragiene, K.~Pyragas, Design of a negative group delay filter via reservoir
  computing approach: Real-time prediction of chaotic signals, Physics Letters
  A 383~(25) (2019) 3088 -- 3094.

\bibitem{Feudel2013}
G.~Ansmann, R.~Karnatak, K.~Lehnertz, U.~Feudel, Extreme events in excitable
  systems and mechanisms of their generation, Phys. Rev. E 88 (2013) 052911.

\bibitem{Feudel2014}
R.~Karnatak, G.~Ansmann, U.~Feudel, K.~Lehnertz, Route to extreme events in
  excitable systems, Phys. Rev. E 90 (2014) 022917.

\bibitem{Ott2013}
H.~L. D.~S. Cavalcante, M.~Ori\'a, D.~Sornette, E.~Ott, D.~J. Gauthier,
  Predictability and suppression of extreme events in a chaotic system, Phys.
  Rev. Lett. 111 (2013) 198701.

\bibitem{Sornette2012}
D.~Sornette, G.~Ouillon, Dragon-kings: Mechanisms, statistical methods and
  empirical evidence, Eur. Phys. J. Spec. Top. 205 (2012) 1--26.

\bibitem{Ott2002}
E.~Ott, Chaos in Dynamical Systems, Cambridge University Press, New York, 2nd.
  ed., 2002.

\bibitem{Gauthier1996}
D.~J. Gauthier, J.~C. Bienfang, Intermittent loss of synchronization in coupled
  chaotic oscillators: Toward a new criterion for high-quality synchronization,
  Phys. Rev. Lett. 77 (1996) 1751--1754.

\bibitem{Venkataramani1996}
S.~C. Venkataramani, B.~R. Hunt, E.~Ott, D.~J. Gauthier, J.~C. Bienfang,
  Transitions to bubbling of chaotic systems, Phys. Rev. Lett. 77 (1996)
  5361--5364.

\bibitem{Appeltant2011}
L.~Appeltant, M.~C. Soriano, G.~Van~der Sande, J.~Danckaert, S.~Massar,
  J.~Dambre, B.~Schrauwen, C.~R. Mirasso, I.~Fischer, Information processing
  using a single dynamical node as complex system, Nature Communications 2
  (2011) 468.

\bibitem{Haynes2015}
N.~D. Haynes, M.~C. Soriano, D.~P. Rosin, I.~Fischer, D.~J. Gauthier, Reservoir
  computing with a single time-delay autonomous boolean node, Phys. Rev. E 91
  (2015) 020801.

\bibitem{Larger2017}
L.~Larger, A.~Bayl\'on-Fuentes, R.~Martinenghi, V.~S. Udaltsov, Y.~K. Chembo,
  M.~Jacquot, High-speed photonic reservoir computing using a time-delay-based
  architecture: Million words per second classification, Phys. Rev. X 7 (2017)
  011015.

\bibitem{Yan2009}
X.~Yan, X.~Su, Linear Regression Analysis: Theory and Computing, World
  Scientific, Singapore, 2009.

\bibitem{abarb91}
H.~D.~I. Abarbanel, R.~Brown, M.~B. Kennel, Variation of lyapunov exponents on
  a strange attractor, J. Nonlinear Sci. Vol. 1: pp. 175-199 (1991).

\bibitem{eck93}
B.~Eckhardt, D.~Yao, Local lyapunov exponents in chaotic systems, Physica D 65
  100-108 (1993).

\bibitem{benet80}
G.~Benettin, L.~Galgani, A.~Giorgilli, J.-M. Strelcyn, Lyapunov characteristic
  exponents for smooth dynamical systems and for hamiltonian systems; a method
  for computing all of them. part 1: Theory, Meccanica 15 (1980) 9--20.

\bibitem{pecora1991}
L.~M. Pecora, T.~L. Carroll, Driving systems with chaotic signals, Phys. Rev. A
  44 (1991) 2374--2383.

\bibitem{pyragas1997}
K.~Pyragas, Conditional lyapunov exponents from time series, Phys. Rev. E 56
  (1997) 5183--5188.

\bibitem{pecora1990}
L.~M. Pecora, T.~L. Carroll, Synchronization in chaotic systems, Phys. Rev.
  Lett. 64 (1990) 821--824.

\bibitem{rulkov1995}
N.~F. Rulkov, M.~M. Sushchik, L.~S. Tsimring, H.~D.~I. Abarbanel, Generalized
  synchronization of chaos in directionally coupled chaotic systems, Phys. Rev.
  E 51 (1995) 980--994.

\bibitem{pyragas1998}
K.~Pyragas, Weak and strong synchronization of chaos, Phys. Rev. E 54 (1996)
  R4508--R4511.

\bibitem{jaeger2001}
H.~Jaeger, Gmd technical report no. 148, Tech. rep., German National Research
  Center for Information Technology (2001).

\end{thebibliography}
\end{document}